 \documentclass[final,5p,times,twocolumn]{elsarticle}

\usepackage{graphicx}
\usepackage{epsfig}
\usepackage{epstopdf}
\usepackage{amssymb}
 \usepackage{lineno}

\journal{Nuclear Physics B}

\begin{document}

\begin{frontmatter}

\title{Vortical field amplification and particle acceleration at rippled shocks}

\author{F. Fraschetti\fnref{fn1}}
\address{Departments of Planetary Sciences and Astronomy, University of Arizona, Tucson, AZ, 85721, USA}

\fntext[fn1]{Associated Member to LUTh, Observatoire de Paris, CNRS-UMR8102 and Universit\'e Paris VII,
5 Place Jules Janssen, F-92195 Meudon C\'edex, France.}

\begin{abstract}
Supernova Remnants (SNRs) shocks are believed to accelerate charged particles and to generate strong turbulence in the post-shock flow. From high-energy observations in the past decade, a magnetic field at SNR shocks largely exceeding the shock-compressed interstellar field has been inferred. We outline how such a field amplification results from a small-scale dynamo process downstream of the shock, providing an explicit expression for the turbulence back-reaction to the fluid whirling. The spatial scale of the $X-$ray rims and the short time-variability can be obtained by using reasonable parameters for the interstellar turbulence. We show that such a vortical field saturation is faster than the acceleration time of the synchrotron emitting energetic electrons. 
\end{abstract}

\begin{keyword}
85-06


\end{keyword}

\end{frontmatter}

 \linenumbers

\section{Introduction}

The origin of cosmic-rays (CRs) still eludes the theoretical and observational efforts in
astroparticle physics since their discovery more than a century ago. Space and
ground-based experiments have been providing us with a wealth of 
multi-wavelength observations to identify the source and investigate the
mechanism of acceleration in various energy bands. Individual shell-type 
Supernova Remnant (SNR) shocks accelerate charged particles and
are believed to provide a significant fraction of the power sustaining 
the observed CR spectrum. Moreover, realistic corrugated shocks travelling in the inhomogeneous 
interstellar space generate turbulence in the compressed post-shock fluid. 

The inhomogeneity of the unshocked ISM observed over several scales \cite{ars95}
is expected to deform the shock surface rippling the initial local planarity 
up to scales many orders of magnitude greater than the thermal ion inertial length.
{\it HST} observations of SN1006 \cite{rksbgs07} constrain the length-scale of the shock ripples to $10^{16} - 10^{17}$ cm.  
We focus on the interaction of a non-relativistic SNR rippled shock 
with the turbulence upstream of the shock, disregarding the contribution of accelerated particles at the shock,
as justified later.

From detection of non-thermal $X$-ray rims \citep{vl03,byuk04},
rapid time-scale variability of $X$-ray hot spots \citep{u07} 
and $\gamma$-ray emission in extended regions \citep{a11}, 
a magnetic field at the shock far exceeding the theoretically 
predicted shock-compressed field has been inferred. 
Whether or not such a magnetic field amplification in SNR is to be associated with 
energetic particles at the shock is still subject of controversy.

Magnetic field amplification might be also relevant to {\it in situ} measurements 
of the plasma downstream of the solar-wind termination shock \citep{b07}, 
where fluctuations have been measured of the same order as the mean,
or to radio observations of Mpc scale shocks at the edge of galaxy clusters \citep{b12}. 
Strong magnetic fields are also required in Gamma-Ray Bursts
(GRB) and Active Galactic Nuclei (AGN) outflows to enable sufficient production of non-thermal radiation. 
In the ISM magnetic energy density and thermal pressure are typically comparable and both amount to a fraction $10^{-9} - 10^{-7} $  of the total internal energy density (including rest mass). Therefore, a compression by an even ultra-relativistic shock (bulk Lorentz factor $\sim 100$), cannot produce the fraction $10^{-3} - 10^{-1} $ predicted by GRB phenomenological models of afterglow light curves \cite{p05}.

The passage of an oblique non-relativistic shock through inhomogeneous medium has been known 
for longtime to generate vorticity in the downstream flow \citep{i64}; in a conducting fluid
the turbulent motion at scale ${\it l}$ with fluid velocity $v_{\it l}$ and local density $\rho$ 
leads exponentially fast to an amplified magnetic field $B^2 = 4\pi\rho v_{\it l}^2$ \citep{ll60}.
The encounter of a shock surface with a density clump, also called 
Richtmyer-Meshkov (RM) instability \cite{b02}, has been also extensively investigated in 
plasma laboratory experiments (see \cite{dr10} and references therein). 

Recent numerical 2D-MHD simulations have shown that 
such an amplification can be very efficient \citep{gj07,i12}.
Ideal MHD applied to 2D rippled shocks has shown that the ISM turbulence 
might amplify exponentially fast the upstream magnetic field 
with a growth rate depending on shock and upstream medium properties \cite{f13}. 
Such an amplification is expected to occur downstream of the blast wave,
regardless the presence of shock-accelerated particles.
Magnetic field may also be enhanced by field line stretching 
due to Rayleigh-Taylor (RT) instability \citep{jns95} at the interface between the ejecta 
and the interstellar medium, i.e., far downstream of the shock.
In contrast with the vortical turbulence, late-time RT turbulence might be affected 
by the highest energy particle gyrating in the downstream fluid
far from the shock \citep{ftbd10}. However, RT structures are unlikely to reach out the blast wave 
(\citep{ftbd10} and references therein) and therefore to interact with vortical turbulence.
Thus the dynamo amplification occurring locally behind the shock can be temporally and spatially disentangled 
from the field line stretching due to RT instability. 

Two-dimensional simulations of relativistic shocks \citep{m11} show that 
small-scale dynamo can operate also downstream of the shocks 
with bulk Lorentz factor of a few unities.
This suggests that the dynamo action downstream of shocks 
might shed light on the energy equipartition at magnetized shocks
of AGN and Gamma-Ray Bursts.

\section{Macroscopic approach to rippled shock}\label{sect_rippled}

\begin{figure}
\includegraphics[width=9cm,height=9cm]{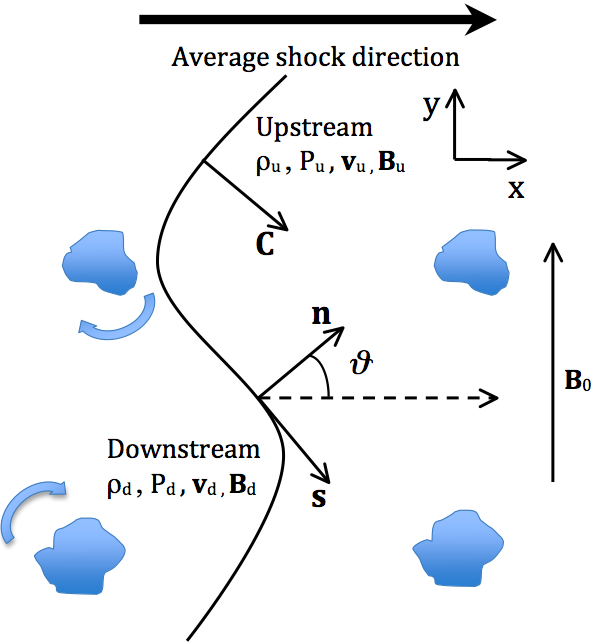}
\caption{Encounter of a shock surface with density enhancement regions: forward and lagging behind regions are formed that generate vorticity in the downstream fluid.}
\label{shock}
\end{figure}

{\it Constitutive equations -} We consider the propagation of a 2D non-relativistic shock front in an inhomogeneous medium. Within the ideal MHD approximation, i.e., with no viscosity or heat conduction, the time evolution of the fluid velocity ${\bf v} = {\bf v} (x,y,t)$ and the magnetic field ${\bf B} = {\bf B} (x,y,t)$, is given, for infinitely conductive fluid, by
\begin{equation}
\left\{
 \begin{array}{cc}
{\partial_t {\bf v}}+({\bf v}\cdot\nabla){\bf v} + \frac{\nabla P}{\rho} + \frac{1}{4\pi\rho} \left[ {\bf B}\times(\nabla \times {\bf B}) \right]  = 0  \\
{\partial_t {\bf B}}=\nabla\times({\bf v}\times{\bf B}) 
\label{induction}
\end{array}
\right.
\end{equation} 
where $\rho$, $P$ are respectively density and hydrodynamic pressure of the fluid (here $\partial_{t} = \partial / \partial t$). 
Note that the current density carried by CRs is here neglected: we aim to identify the growth of the magnetic energy as generated by the vortical motion of the background fluid only. Plasma heating by the shock might reduce the energy deposited in the magnetic turbulence 
and will be considered in a forthcoming publication.

{\it Vorticity downstream of MHD shock -} The vorticity shock-generated is transported along the flow ``frozen'' into the fluid in the inviscid approximation (Helmholtz-Kelvin theorem). The medium upstream of the shock has ${\omega} = 0$. The vorticity is calculated downstream at a distance from the shock large enough that the shock is infinitely thin, i.e., the thickness of the shock is much smaller than the local curvature radius at every point of the shock surface. 

At a rippled shock the MHD Rankine-Hugoniot jump conditions cannot be applied globally as the directions normal and tangential vary along the shock surface. For a 2D shock propagating at average in the direction $x$ (all quantities are independent on $z$, see Fig.\ref{shock}), from the velocity field of the flow ${\bf v} = (v_x, v_y, 0)$, the vorticity is given by $|{\omega}| = |\nabla \times {\bf v}| =  \omega_z$.  
We use a local natural coordinate system $(\hat n,\hat s)$, where $\hat n = ({\rm cos} \vartheta(t, s), {\rm sin} \vartheta (t, s))$ is the coordinate along the normal to the shock surface, $\hat s = ({\rm sin} \vartheta (t, s), {\rm -cos} \vartheta (t, s))$ is the coordinate parallel to the shock surface (Fig.\ref{shock}). 
We consider a seed-magnetic field upstream uniform and normal to the average direction of motion (${\bf B}_0 = (0, B_0^y, 0)$, or $B_n=~B_0 {\rm sin} \vartheta$ and $B_s=-B_0 {\rm cos} \vartheta$, see Fig.\ref{shock}). 

The turbulent field is assumed to be much greater than the shock-compressed field in the downstream flow, in agreement with observations, so that the amplification is efficient at the smallest scales (see Sect. \ref{OBS}).  
Thus, the vorticity produced downstream of a 2D shock propagating in an inhomogeneous medium with a uniform perpendicular upstream magnetic field (same as for parallel shock \cite{f13}) can be recast, neglecting obliqueness, in a simple form (we use $\partial_{x_i} = \partial/\partial_{x_i}$):
\begin{equation}
|\delta \omega_z|  =  \frac{r-1}{r} \left[ \left(\frac{C_r}{\rho}\right)_u \partial_s \rho + \partial_s C_r \right]  -  \frac{B_n \delta B_s}{4\pi \rho C_r} \partial_s \vartheta ,
\label{dom_perp}
\end{equation}
where $r=\rho_d / \rho_u$ is the compression ratio at the shock, $C_r$ is the shock speed relative to the upstream frame, $\delta B_s$ is the jump across the shock of the magnetic field in the direction locally tangential to the shock surface including the Rankine-Hugoniot compressed seed field and the turbulently amplified field and $B_n$ is the component in the direction locally normal to the shock surface including the unchanged Rankine-Hugoniot and the turbulent components. 

\begin{figure}
\includegraphics[width=8cm,height=8cm]{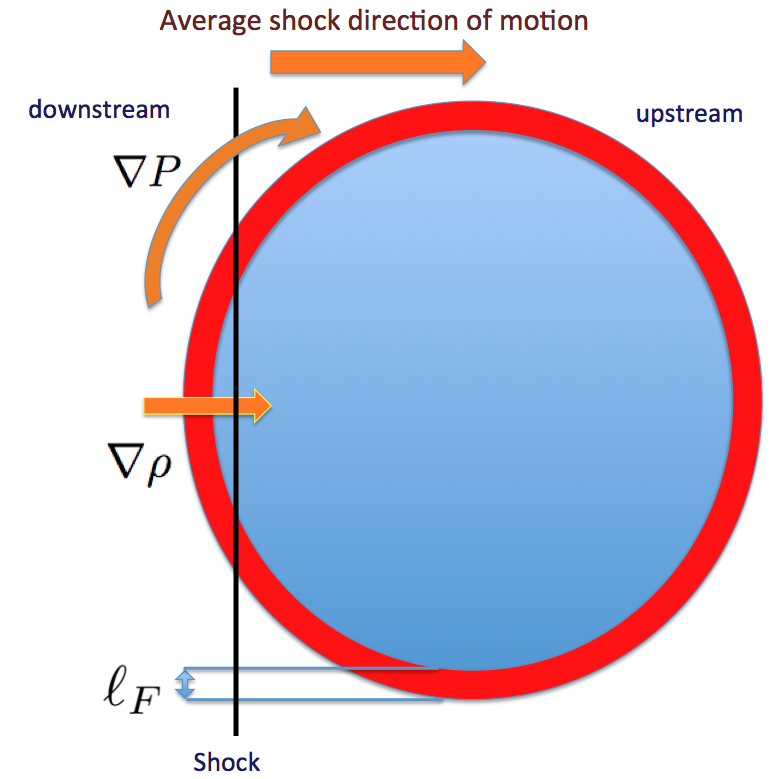}
\caption{Generation of the baroclinic term of the vorticity at the shock crossing in the condensation layer of thickness $\ell_F$. } 
\label{shock2}
\end{figure}

{\it Turbulent field amplification -} The vortical turbulence described in the previous sub-section exponentially amplifies  the total magnetic field. Since the amplification time-scale is of the order of the smallest eddies turnover time \cite{bk05}, the saturation occurs much faster at small-scale \cite{k05}. This is the key feature of the small-scale dynamo. 
The unperturbed field is initially too weak to affect the fluid velocity field and the turbulent field grows exponentially fast, until the magnetic energy produces non-negligible effects on the velocity field and then saturates.

The small-scale dynamo theory predicts that the turbulent field obeys an unbounded exponential amplification at a rate $\beta$ \citep{k05,ka12}: $d\varepsilon /dt = 2\beta \varepsilon$, where $\varepsilon = B^2 / 8\pi \rho$ is the total magnetic energy per unit of mass, including seed and turbulent fields. As shown in \cite{k05}, the isotropy and homogeneity of the fluid velocity correlation entails the following simple relation between the amplification rate of $\varepsilon$ and the vorticity generated downstream of the shock: $\beta \simeq (\pi/3) \delta \omega_z$. 

If we recast Eq.(\ref{dom_perp}) as $ |\delta \omega_z| = (3/\pi) (\tau^{-1} - \alpha \varepsilon)$, then $\varepsilon$ satisfies 
\begin{equation}
\frac{d \varepsilon}{dt} = 2 (\tau^{-1} - \alpha\varepsilon)\varepsilon
\label{epsilonEq}
\end{equation}
where $\tau^{-1} = \frac{\pi}{3} \frac{r-1}{r} \left[ (C_r/\rho)_u \partial_s \rho + \partial_s C_r \right] $ is the local growth rate of $\varepsilon$ and $\alpha =  (2 \pi/3) \partial_s \vartheta /C_r $ is the local back-reaction; the initial condition for Eq. (\ref{epsilonEq}) is $\varepsilon (0) = \varepsilon_0 = v_A^2 / 2= B_0^2/8\pi\rho $. In Eq.(\ref{epsilonEq}) we have assumed that the turbulence dominates over $B_0$, i.e., $\delta B_s/\sqrt{8\pi \rho} \sim \sqrt{\varepsilon}$ and $ B_n/\sqrt{8\pi \rho} \sim \sqrt{\varepsilon}$: the turbulence grows isotropically downstream at the shock curvature scale as a consequence of the isotropy of the flow velocity field \citep{k05}. 

Neglecting the time dependence of $\tau$ (the magnetic modes grow slowly for initially weak field \citep{k05}), the solution is readily found: 
\begin{equation}
\frac{\varepsilon}{ \varepsilon_0} (t) = \left(\frac{B}{B_0}\right)^2 (t) = \frac{e^{2t/\tau}}{1-\alpha\tau(1-e^{2t/\tau})v_A^2 /2},
\label{exact}
\end{equation}
for a uniform average interstellar matter density.

\section{Comparison with multiwavelength SNR observations}\label{OBS}

\begin{figure}
\includegraphics[width=9cm,height=5.5cm]{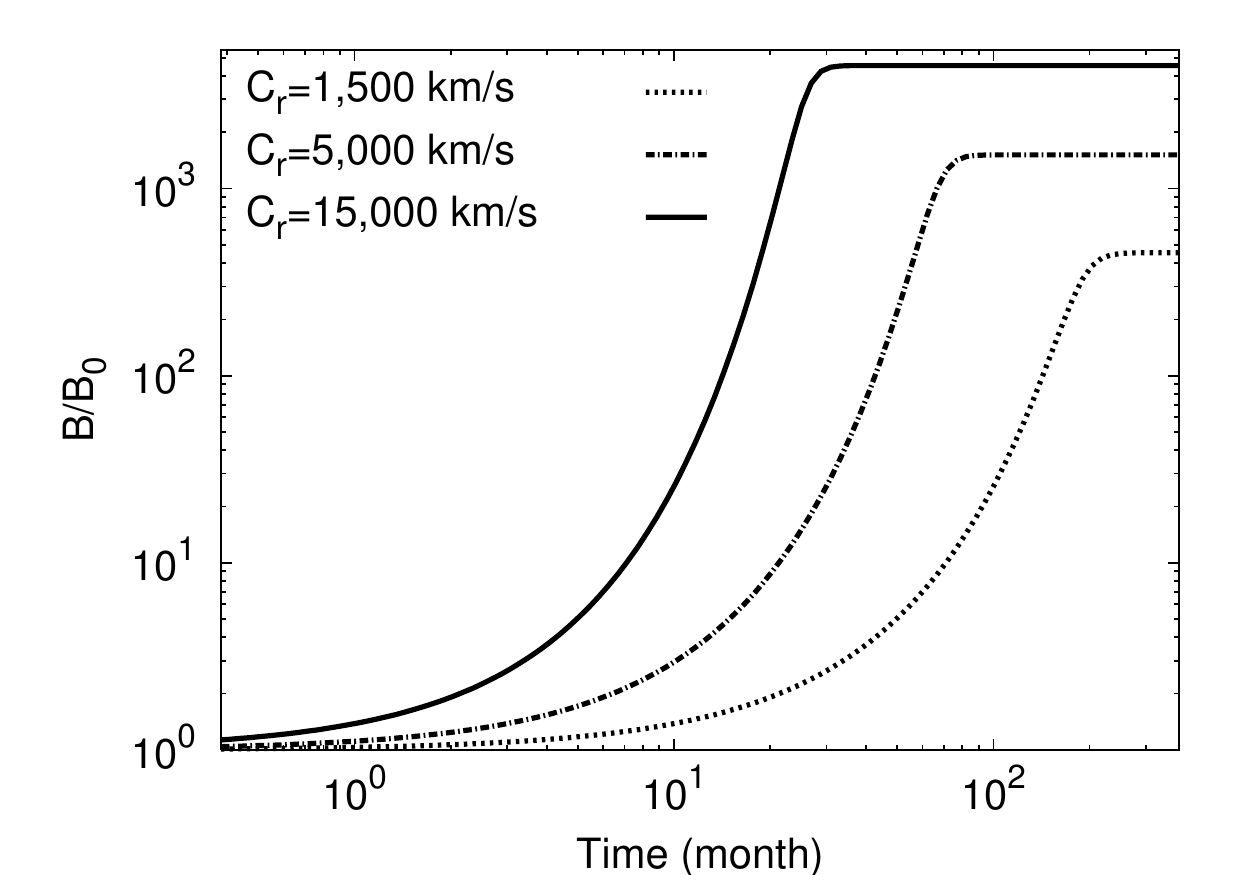}
\caption{Saturation of the total magnetic field for various shock speed $C_r$ is shown: $C_r = 1,500$ km$/$s, $C_r = 5,000$ km$/$s, $C_r = 15,000$ km$/$s, assuming $R_c = 10^{17}$ cm and $\ell_F = 10^{16}$ cm, that results in $\tau \lesssim \ell_F/C_r \sim 3$ years for $C_r \gtrsim 5,000$ km$/$s ($\vartheta = 0.1$ rad, $r=4$ and $v_A = 10^{-4} c$).}
\label{Cr}
\end{figure}

The growth rate of $\delta B$ can be approximated as $\tau^{-1} \sim {C_r}(R_c +\ell_F)/(R_c \ell_F)$, where $R_c$ is the local curvature radius of the shock surface. Thus $\tau^{-1}$ increases with shock speed and it depends mainly on hydrodynamic quantities. If $\ell_F \ll R_c$, it holds $\tau \sim \ell_F/C_r$: the amplification saturates faster for smaller $\ell_F$.

As the magnetic field strengthens, it reacts to field lines whirling halting the turbulence growth. In more general terms, as the field increases by dynamo action it also releases its tension by unwinding at a rate of order of Alfv\'en speed:  the backreaction grows with the turbulent field Alfv\'en speed \citep{k05}. The local back-reaction of the field $\alpha \sim \partial_s \vartheta /C_r $ can be estimated by $\alpha \sim \vartheta/(R_c C_r) $.

Fig.\ref{Cr} depicts the growth of the turbulent field for various shock speeds, assumed constant in time: given an ISM field of the order of $B_0 \sim 3 \mu $G, the turbulent field saturates at $B \sim 1.2-3.$ mG for $C_r = 1,500 - 5,000$ km$/$s on the year time-scale. Such a rapid growth of magnetic energy is compatible with $X$-ray observations of SNRs $RX J1713.7-3946$ ($C_r < 4,500$ km$/$s \citep{u07}) and Cas A \citep{pf09} brightness variations detected on year time-scale in small-scale hot spots structures, attributed to synchrotron electron cooling. Using $R_c = 10^{17}$ cm and $\ell_F = 10^{16}$ cm, we find an amplification to $B \sim 3. $ mG within $3$ years. Such a value of $\ell_F$ is to be compared with the spatial scale of the {\it Chandra} $RX J1713.7-3946$ bright spots, estimated as $\lesssim 0,03$ pc. 
Similar length ($\sim 10^{14} - 10^{16}$ cm) and time ($\sim 1$ yr) scales are found in simulations of the effects of magnetic field turbulence on the observed synchrotron emission images and spectra in SNRs \citep{bue08}. Thus, the magnetic energy increase and the $X$-ray variability might have a time-scale ($\sim 1$ yr) much lower than the SNR hydrodynamic time-scale and might occur in middle-aged, not necessarily young, SNRs ($RX J1713.7-3946$ age is estimated as $1,600$ yr \citep{sg02}). The high shock speed $C_r \sim 15,000$ km$/$s in Fig.\ref{Cr} is comparable to observations of the youngest SNR in our galaxy, i.e., $100$ years old G$1.9+0.3$ \citep{rbghhp08}. Thus, a rapid field saturation even up to $B \sim 10$ mG is predicted at SNR shocks within a few months.

\section{Constraints on particle acceleration} 

If the thickness of the density steepening layer at the boundary of the ISM density clumps is identified as the Field length $\ell_F$, for typical cold ISM, we can use $\ell_F \simeq 3.3 \times 10^{16}$ cm, with an uncertainty depending on ionization and heating$/$cooling properties \cite{bm90}. Such an $\ell_F$ is compatible with the ripple scale inferred by optical observations \cite{rksbgs07}. Thus, for a typical middle-aged SNR with shock speed $C_r \sim 5,000$ km$/$s, the growth time-scale of the vortical turbulence is $\tau \simeq \ell_F/C_r \simeq 6.7 \times 10^7$ s $\sim 1.9$ yrs. 

A simple argument shows that $\tau$ is shorter than the typical acceleration time-scale for energetic electrons at the shock, i.e., $\tau_{acc}$. Modulo a factor of order of unity, $\tau_{acc} \simeq \kappa_E / C_r^2$, where the diffusion coefficient $\kappa_E$ (neglecting its change across the shock) depends on the particle energy and on the magnetic field orientation. If the seed magnetic field is parallel to the local shock normal, the diffusion coefficient governing the  electron acceleration $\kappa_E$ is necessarily greater than the Bohm diffusion coefficient $\kappa_B$, corresponding to $\lambda \simeq r_g$, where $\lambda$ is the mean free path of the charged particle and $r_g = pc/eB$ is the particle gyroradius.  The typical energy of an electron emitting synchrotron radiation at $5$ keV in an amplified magnetic field $B \sim 100  \mu$G is $E \sim 50$ TeV. Thus, for an energetic  electron diffusing at the shock in the Bohm regime, $\kappa_E = r_g c/3 \simeq 3.3 \times 10^{23} E_{13}/B_{2}$ cm$^2 /$s, where $E_{13}$ is the electron energy in units of $10$ TeV and $B_{2}$ the magnetic field in the $X-$ray rim in units of $100 \mu$G. Thus, we obtain $\tau \simeq \kappa_E/C_r^2 = 1.9$ years $\simeq \tau$. 

Bohm diffusion, despite largely used in the literature because of the lack of self-consistent diffusion theory in strong turbulence, describes transport only for a very limited range of particle energy (see \cite{clp02,fg12}). Since the scattering diffusion coefficient $\kappa_E$ in most cases is much greater than $\kappa_B$, the inferred field amplification might occur on a time-scale much shorter than acceleration time-scale of particles scattering back and forth across the shock. Our simple estimate, derived from the $X-$ray synchrotron parameters and the inferred strong field, holds regardless the location of the emitting region, whether upstream or downstream of the shock. The change of the structure of the turbulence across the shock, due to the anisotropic shock-compression and the vortical amplification downstream shown here, is not expected to modify significantly our estimate of $\tau_{acc}$.

\section{Conclusion} \label{sect_concl}

By applying first principles to a 2D rippled shock, we have outlined the derivation of  temporal evolution and saturation of the turbulent magnetic field downstream of the shock, including the non-linear field back-reaction. We conclude that the saturation of $B$ by small-scale dynamo action depends on the shock speed, on the thickness of the density steepening layer at the boundary of the ISM density clumps and on the shock curvature radius, but not on the size of the ISM clumps. Our finding shows that small-scale dynamo might explain non-thermal $X$-ray observations and agrees with the optical upper limit on the scale of shock ripples. The magnetic field enhancement described here occurs faster than acceleration time-scale of synchrotron emitting energetic electrons.

{\it Acknowledgment -- } The author thanks ISSI for providing stimulating environment. 
The support from NASA through the Grants NNX10AF24G and NNX11AO64G is gratefully acknowledged.

\label{}

\end{document}